\documentclass[aps,preprint,showpacs]{revtex4}

\usepackage{graphicx}
\usepackage{epsfig}
\usepackage{amssymb}
\usepackage{amsmath}
\usepackage{psfrag}

\newcommand{\rg}{{\bf r}}
\newcommand{\Eg}{{\bf E}}
\newcommand{\Gg}{{\bf G}}
\newcommand{\pg}{{\bf p}}

\newcommand{\ddd}{{\mathrm{d}}^2}

\newcommand{\Imag}{\mathrm{Im} \, }

\begin{document}

\title{Fluctuations of the local density of states probe localized surface plasmons on disordered metal films}
\author{V. Krachmalnicoff, E. Castani\'e, Y. De Wilde and R. Carminati}
\email{remi.carminati@espci.fr}
\affiliation{Institut Langevin, ESPCI ParisTech, CNRS, 10 rue Vauquelin,
75231 Paris Cedex 05, France}

\begin{abstract}
We measure the statistical distribution of the local density of optical states (LDOS) on disordered semi-continuous metal films. 
We show that LDOS fluctuations exhibit a maximum in a regime where fractal clusters dominate the film surface. 
These large fluctuations are a signature of surface-plasmon localization on the nanometer scale.
\end{abstract}


\maketitle


The study of light scattering and transport in disordered media is stimulated by fundamental issues
in mesoscopic physics~\cite{ShengBook,AkkermansBook}, including Anderson localization~\cite{John},
and by the development of imaging techniques in complex media~\cite{SebbahBook}. Moreover, disordered dielectrics
and metal-dielectric composites provide new classes of photonic materials for the control of light propagation 
and the enhancement of light-matter interaction~\cite{LakhtakiaBook,ShalaevBook,phot_glasses,Wiersma2008,SLM,Lodahl2010}. 
Disordered semi-continuous metallic films are a particularly striking example, since they exhibit optical properties 
that strongly differ from those of bulk metals and ensembles of isolated nanoparticles~\cite{ShalaevBook}. 
In these systems, the interplay between intrinsic material excitations - surface plasmons - 
and random scattering by multiscale (fractal) metallic clusters leads to spatial localization of the electromagnetic field in subwavelength areas 
(hot spots)~\cite{Stockman1997,Gresillon1999,Shalaev_review}. At a given frequency, surface-plasmon modes consist
of one or several hot spots, and can be localized (i.e. insensitive to the sample boundaries) or delocalized (spread over the entire system).
The coexistence of both types of modes results from the self-similarity of the structure, and is referred to as inhomogeneous 
localization~\cite{Stockman1997,Stockman2001,Cao2006}.

In this Letter, we study experimentally the behavior of the local density of optical states (LDOS) on disordered fractal metallic films.
The LDOS is a fundamental quantity for the characterization of the optical properties of complex systems.
It drives the spontaneous emission of light by dipole emitters~\cite{Sipe1984},
and is also connected to macroscopic transport properties~\cite{Lagen1996,Mirlin2000,Carminati2009,Froufe2009}.
In a disordered medium, changes in the LDOS probe the local environment~\cite{Luis0708}, 
the photon transport regime~\cite{Beenakker2002,Pierrat2010}, or drive long-range correlations of speckle patterns~\cite{Skipetrov2006}. 
Recent experiments have provided evidences of changes in the statistical distribution of the LDOS in disordered dielectric 
media~\cite{Vahid2010,Mosk2010}, but in regimes in which light localization is not observed. Disordered metallic films offer
the advantage to generate localized surface-plasmon modes, and to exhibit sharp changes of their optical properties when multiscale
clusters appear, thus being good candidates for the observation of substantial changes of the LDOS. In particular, since the LDOS
fluctuations are expected to be sensitive to changes of the structure of electromagnetic modes, one can expect to get a signature
of inhomogeneous localization in the LDOS statistics. Moreover, although the intense field in hot spots
has been used for the enhancement of non-linear response of optical surfaces~\cite{ShalaevBook}, 
little is known about the possibility of using these
localized plasmon modes for quantum-electrodynamics studies~\cite{Lodahl2010}.
Here we provide the first measurements of the LDOS on disordered metallic films, using nanoscale fluorescent emitters. We show
that the LDOS exhibits large fluctuations at the threshold corresponding to the appearance of fractal clusters. We interpret the
measurements in terms of the inverse participation ratio, a parameter that measures the spatial extent of eigenfunctions~\cite{Mirlin2000}, 
and we show that the large fluctuations are a signature of localized surface-plasmon modes.

\begin{figure}[h]
\begin{center}
\includegraphics[width=7cm]{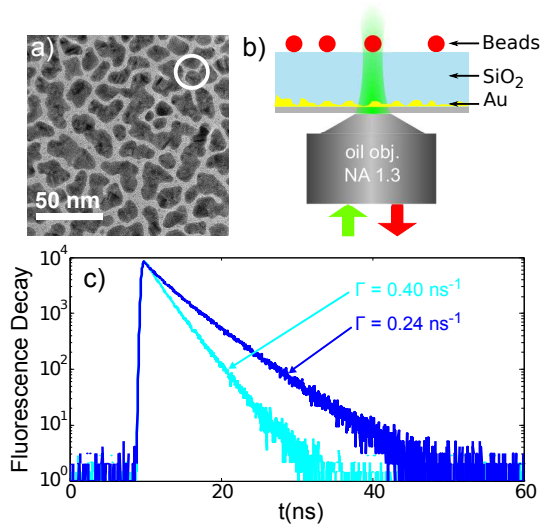}
\caption{\label{fig:geometry} (Color online) (a): TEM image of a sample with surface fraction $f=67 \%$.
The white circle indicates the size of a fluorescent bead. 
(b): Scheme of the experimental configuration. Excitation and detection are performed with a confocal microscope
on a single fluorescent bead. A $40$ nm thick  silica spacer
is deposited on each sample. (c): Typical decay curves recorded on two different beads on the sample in panel (a).}
\end{center}
\end{figure}

A sketch of the experimental setup and a typical sample are shown in Fig.~\ref{fig:geometry}.
To fabricate the samples, gold was deposited under high vacuum
conditions by electron beam vapor deposition (deposit rate $\simeq 1$~\AA/s) on a glass microscope
cover slide previously covered by a very thin (5~nm) SiO$_2$ layer.
Thicknesses were monitored
by a quartz crystal microbalance. A 40~nm SiO$_2$ spacer was
then deposited on the gold layer and fluorescent polystyrene beads
(Invitrogen Fluospheres Red 580/605, diameter 25~nm) were spin-coated
on top of the sample (final beads density $\simeq
10^{-2}\mathrm{~\mu m}^{-2}$). The thickness of the SiO$_2$ spacer was
chosen to be large enough to avoid fluorescence quenching but as small
as possible in order to reduce averaging over several gold clusters. 
Eight samples, with gold mass thickness
ranging from 1 (isolated gold clusters, gold filling fraction $f=30\%$) to 20~nm 
(almost continuous gold film, $f=99\%$), were prepared together with samples with a thinner (5~nm)
SiO$_2$ top layer deposited on carbon-copper grids for TEM
observation. TEM images were used to measure the gold filling fraction
and to characterize the topology of the sample. 
A typical TEM image of a 5~nm thick gold film is
shown in Fig.~\ref{fig:geometry} (a). For
fluorescence lifetime measurements the beads are individually excited through an
oil immersion objective (x100, NA 1.3) mounted in an inverted confocal
microscope (Olympus IX-71). The excitation is performed at 560~nm
with a mode-locked super-continuum laser
(Fianium SC450) that provides $\simeq 100$~ps pulses at a repetition
rate of 10~MHz. The excitation intensity is about $10^3$~W/cm$^2$.
Fluorescence photons are collected through the same objective and are
separated from the excitation light using a dichroic mirror followed by a
set of filters centered at 607~nm with a bandwidth of 70~nm. 
Time-resolved photon detection is provided by combining
a single-photon avalanche photodiode (Perkin Elmer SPCM-AQR-15) and a
time correlated photon counting system (PicoQuant TimeHarp 200). Two
typical decay curves are shown in Fig.\ref{fig:geometry} (c). A
strictly mono-exponential decay is observed. For all statistics presented in this work,
we have measured the decay rate of approximatively 100 beads on each sample.
%


Measuring the lifetime of the excited state $\tau$ amounts to measuring the spontaneous decay rate $\Gamma = 1/\tau$. 
For a dipole emitter, the latter is $\Gamma = \pi \omega/(\hbar \epsilon_0) |\pg |^2 \rho(\rg,\omega)$, where $\rho(\rg,\omega)$ 
is the LDOS, $\omega$ is the emission frequency and $\pg$ the transition dipole~\cite{Sipe1984}.
The LDOS describes the electromagnetic environment, and can be computed
using the dyadic Green function that describes the response at point $\rg$ to a point electric dipole
${\bf p}$ located at point $\rg^\prime$ through the relation $\Eg(\rg) = \mu_0 \, \omega^2 \, \Gg(\rg,\rg^\prime) {\bf p}$. 
It reads $\rho(\omega,\rg) = 2 \omega/(\pi c^2) \, \Imag \mathrm{Tr} \, \Gg(\rg,\rg)$,
where $\mathrm{Tr}$ denotes the trace of a tensor. The trace in this expression corresponds to an LDOS
averaged over the transition dipole orientation, which is the quantity of interest in our measurements since the
beads contain many molecules with arbitrary orientations.

\begin{figure}[h]
\begin{center}
\includegraphics[width=7cm]{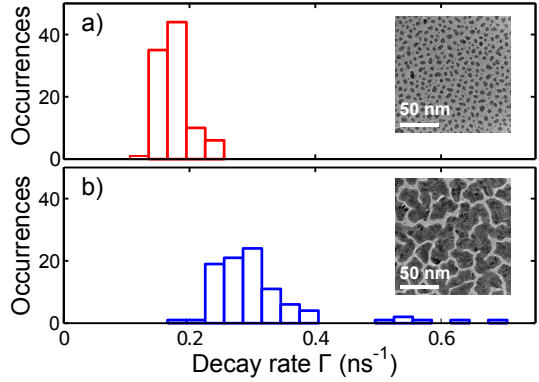}
\caption{\label{fig:two_films}(Color online) Decay rate distributions for two different samples. (a): $f=30\%$ (isolated nanoparticles). 
(b): $f =82\%$ (fractal clusters). Insets: TEM image of each sample.}
\end{center}
\end{figure}

The disordered metal films are  made of clusters of gold nanoparticles. When the gold filling fraction $f$ increases, three typical regimes
are identified. (1)~At low $f$, the sample consists in isolated gold nanoparticles (or small islands) on a 
dielectric substrate (glass). (2)~When $f$ increases, one enters a regime with growing clusters exhibiting self-similarity (fractals), that are 
responsible for peculiar optical properties (e.g. anomalous spectral absorption)~\cite{Shalaev_review}. 
Electrical percolation also occurs in this regime. (3)~Finally, the semi-continuous structure evolves toward an almost continuous metal film
filled with dielectric voids. The decay rate distributions obtained for two films corresponding to regimes (1) and (2) are shown in 
Fig.~\ref{fig:two_films}, together with the corresponding TEM images. 
The difference between both distributions is substantial. From regime (1) in Fig.~\ref{fig:two_films}(a) (isolated nanoparticles)  
to regime (2) Fig.~\ref{fig:two_films}(b) (fractal clusters), the distribution broadens and the averaged value increases. 
One also observes a long tail Fig.~\ref{fig:two_films}(b), corresponding to a few beads displaying a large increase of the decay rate. 
Such large values are only observed in this regime ($f \gtrsim 80\%$). 
Changes of the spontaneous decay rate on these structures is attributed to changes in the LDOS,
reflecting changes of the structure of the electromagnetic modes.

The behaviors in Fig.~\ref{fig:two_films} (a) and (b) can be understood qualitatively,
based on previous studies of the near field on such films. 
Under far-field excitation, the fluctuations of the near field at the film surface in regime (1) generate a speckle 
pattern that can be understood in usual terms, without invoking the collective resonant interactions giving rise to the 
hot-spots structures~\cite{Cao2005,Laverdant2008}.
In regime (2), the near-field exhibits giant fluctuations (hot spots) which are spatially localized much below the illumination
wavelength, on a scale of a few tens of nanometers~\cite{Gresillon1999}. The strong spatial fluctuations of the amplitude of
each mode explain the enhanced fluctuations of the LDOS in this regime.

\begin{figure}[h]
\begin{center}
\includegraphics[width=7cm]{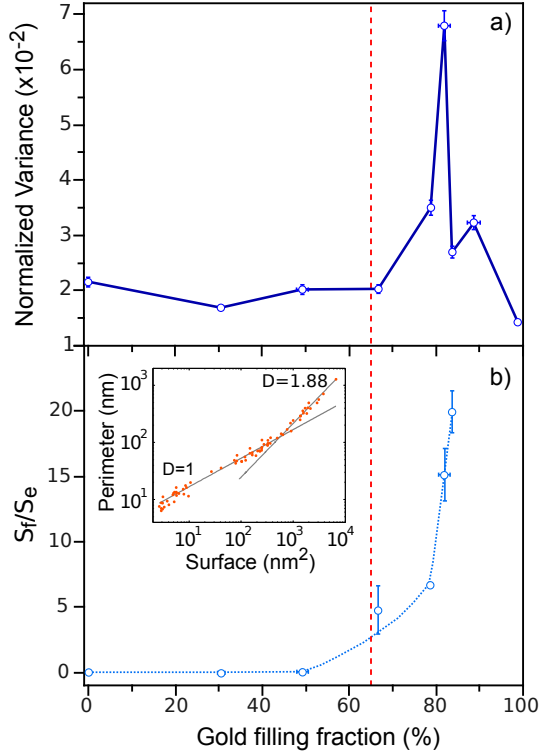}
\caption{\label{fig:variance} (Color online) (a): Normalized variance $\sigma^2(\Gamma)/\langle \Gamma \rangle ^2$ versus the gold filling fraction $f$.
A sharp peak is visible for $f = 82 \%$ and is the signature of surface-plasmon modes localization.
(b): Fractal parameter $S_f/S_e$ versus $f$. The dotted line is a guide for the eye.
Inset: Perimeter versus surface of the clusters of a given sample ($f=79 \%$) deduced from a TEM
image (double logarithmic scale). The solid lines give the fractal dimension $D$ showing the existence of euclidian ($D=1$)
and fractal clusters ($D=1.88$).}
\end{center}
\end{figure}

A feature of the hot-spots regime is the existence of spatially localized (surface-plasmon) modes, together with
delocalized modes (inhomogeneous localization). The delocalized modes are formed by a number of separated hot spots,
that are distributed in a region with size comparable to the sample size. The localized modes are insensitive to the sample 
boundaries, as in Anderson localization~\cite{Stockman2001}. Motivated by the observation in Fig.~\ref{fig:two_films}, we have 
carried out measurements of statistical distributions of decay rates on samples with different filling fractions $f$, covering the three
regimes discussed above. 
As a measure of the fluctuations, we have computed the normalized variance 
$\sigma^2(\Gamma)/\langle \Gamma \rangle^2 = \langle \Gamma^2 \rangle/\langle \Gamma \rangle^2 -1$. The result is shown in
Fig.~\ref{fig:variance}(a). 
In the region corresponding to $f > 65\%$, we observe an asymmetric double-peak structure, 
with a sharp maximum of the decay rate (LDOS) fluctuations at $f = 82\%$.
Since these enhanced fluctuations were expected in the regime where fractal clusters
play a role, we have analyzed the TEM images of each sample in order to compute the ratio $S_f/S_e$ of the surface occupied by
fractal clusters and the surface occupied by euclidian clusters. 
On a given sample, fractal and euclidian clusters were identified through
their dimension $D$ such that $P \propto \Sigma^{D/2}$, where $\Sigma$ is the surface and $P$ the perimeter of a given cluster.
In semi-continuous metal films euclidian and fractal clusters are known to have $D=1$ and $D=1.88$, respectively.
For the ensemble of clusters of a given sample, one obtains a behavior as that shown in the inset of Fig.~\ref{fig:variance}(b),
that allows us to attribute an euclidian or fractal character to each cluster.
The ratio $S_f/S_e$ is taken as a measure of the fractal character of a given sample.
Its evolution with the filling fraction $f$ is shown in Fig.~\ref{fig:variance}(b). We observe a sharp increase for $f > 65 \%$ that characterizes 
the appearance of fractal clusters (indicated by the vertical red dotted line in Fig.~\ref{fig:variance}). The
maximum of LDOS fluctuations is observed for $f = 82 \%$, i.e. in a regime where fractal clusters dominate ($S_f/S_e \sim 10$),
and in which localized surface-plasmon modes are expected according to the inhomogeneous localization concept.
The correlation between enhanced LDOS fluctuations and the existence of localized modes is the main result of this Letter. 
As we shall see, this correlation is supported by an analysis based on the inverse participation ratio.


The inverse participation ratio $R_{IP}$ is a quantity used in the theory of Anderson localization to measure the spatial extent 
of wavefunctions, and to determine the localization transition~\cite{Mirlin2000}. It is defined by
\begin{equation}
R_{IP} = \frac{\int |\Eg(\rg)|^4 \, \ddd r}{(\int |\Eg(\rg)|^2 \, \ddd r)^2}
\end{equation}
where in our case the integral is performed along a plane parallel to the sample surface.
For spatially localized modes, $R_{IP}$ is independent on the sample size $L$, whereas for extended modes, 
$R_{IP}$ scales as $L^{-2}$~\cite{Brouers2000}. 
In our case, $R_{IP}$ can be used to measure the surface occupied by hot-spot modes (localized or delocalized). 

In order to connect the LDOS fluctuations to $R_{IP}$, we can make the following hypothesis: At 
a given point $\rg$ and at a given frequency $\omega$, the electric field is dominated by one mode.
This means that the probability of having more than one mode giving a high electric field at a given point and 
given frequency is very small. In this case, the LDOS is essentially~\cite{NovotnyBook}:
\begin{equation}
\rho(\omega,\rg) \propto \sum_n |\Eg_n(\rg)|^2 \, \delta(\omega-\omega_n) \simeq \frac{1}{\Delta \omega} |\Eg(\rg)|^2
\label{eq:approxLDOS}
\end{equation}
where $\Delta \omega$ is the spectral width of the mode and $|\Eg(\rg)|^2=I(\rg)$ its local intensity.
Under this approximation, the inverse participation ratio reads:
\begin{equation}
R_{IP} = \frac{\int |I(\rg)|^2 \, \ddd r}{(\int I(\rg) \, \ddd r)^2} = \frac{\langle \rho^2 \rangle}{S \, \langle \rho \rangle^2}
\end{equation}
where $S$ is the sample surface. In the last equality, we have assumed ergodicity so that spatial and
statistical averaging have been considered as equal~\cite{Shalaev2003}. This expression shows that measuring
LDOS fluctuations provides a direct measurement of $R_{IP}$. As a result, we can infer the increase of  LDOS fluctuations
as a signature of an increased contribution of localized modes. Indeed, for delocalized modes, one has $N$ hot spots 
($N \gg 1$), each of them with typical extent $\xi$~\cite{Stockman2001}, and the inverse participation is $R_{IP} \sim (N \xi^2)^{-1}$.
For localized modes, one has $N\sim 1$, with a localization length $\xi_{l} \lesssim \xi$, so that $R_{IP} \sim \xi_l^{-2} \gg (N \xi^2)^{-1}$.
This simple analysis shows that the peak in the LDOS fluctuations observed in Fig.~\ref{fig:variance}(a) is the signature
of an increased number of localized surface-plasmon modes in the regime where the disordered film contains a substantial fraction
of fractal clusters.

The approximation leading to Eq.~(\ref{eq:approxLDOS}) is supported by the similarity between the 
double-peak structure of the relative variance observed in Fig.~\ref{fig:variance}(a) and the relative variance of
the near-field intensity pattern measured by scanning near-field optical microscopy (SNOM)
on similar samples~\cite{Cao2006}. These SNOM measurements have led to the conclusion that localized modes 
should dominate around the percolation threshold (but not exactly at percolation), in agreement from expectations 
resulting from numerical simulations~\cite{Shalaev2003}.
An advantage of our direct measurement of the LDOS is that the local intensity of both radiative and non-radiative (dark) 
modes is probed with the same weight, and with a well-defined instrumental response function. 
Even in a SNOM experiment that is able to probe the intensity of both
types of modes, scattering is necessary in order to couple external radiation to non-radiative modes, whereas
radiative modes can be directly excited. This asymmetry is avoided by direct LDOS measurements.


In conclusion, we have measured the statistical distribution of the spontaneous decay rate of fluorescent nanosources on 
disordered fractal metal films. Changes in the decay rates have been attributed to changes in the LDOS, thus providing
the first analysis of LDOS fluctuations on such systems.  We have found that at the onset of the existence of fractal clusters, 
the normalized fluctuations of the LDOS exhibit a sharp maximum. This maximum is attributed to
the presence of localized surface-plasmon modes, and coincides with a maximum of the inverse participation ratio. 
These experiments show that LDOS fluctuations carry sensitive signatures of photon transport in complex media,
and seem to confirm the concept of inhomogeneous localization on self-similar disordered structures. They also show
that localized plasmon modes can substantially modify the LDOS, on a scale in the range of tenth of nanometers, thus
providing an alternative to dielectric microcavities and plasmonic nanoantennas for studies of light-matter interaction
at the single-emitter level.

We acknowledge S. Bidault, C. Boccara, E. Fort and S. Gr\'esillon for illuminating discussions, and K. Balaa and X. Xu
for their help in the sample preparation and characterization. 
This work was supported by the EU Project {\it Nanomagma} NMP3-SL-2008-214107 and by CNANO Ile de France.
V.K. and E.C. acknowledge grants from the Fondation PGG and the French DGA.


\end{document}